# Varying the Population Size of Artificial Foraging Swarms on Time Varying Landscapes

Carlos Fernandes[1,3], Vitorino Ramos[2], Agostinho C. Rosa[1]

[1] LaSEEB-ISR-IST, Technical Univ. of Lisbon (IST),
Av. Rovisco Pais, 1, TN 6.21, 1049-001, Lisbon, PORTUGAL
{cfernandes,acrosa}@laseeb.org
[2] CVRM-IST, Technical Univ. of Lisbon (IST),
Av. Rovisco Pais, 1, 1049-001, Lisbon, PORTUGAL
vitorino.ramos@alfa.ist.utl.pt
[3] EST-IPS, Setúbal Polytechnic Institute (IPS),
R. Vale de Chaves - Estefanilha, 2810, Setúbal, PORTUGAL

**Abstract.** Swarm Intelligence (SI) is the property of a system whereby the collective behaviors of (unsophisticated) entities interacting locally with their environment cause coherent functional global patterns to emerge. SI provides a basis with which it is possible to explore collective (or distributed) problem solving without centralized control or the provision of a global model. In this paper we present a Swarm Search Algorithm with varying population of agents. The swarm is based on a previous model with fixed population which proved its effectiveness on several computation problems. We will show that the variation of the population size provides the swarm with mechanisms that improves its self-adaptability and causes the emergence of a more robust self-organized behavior, resulting in a higher efficiency on searching peaks and valleys over dynamic search landscapes represented here – for the purpose of different experiments – by several three-dimensional mathematical functions that suddenly change over time. We will also show that the present swarm, for each function, self-adapts towards an optimal population size, thus self-regulating.

## 1 Introduction

Swarm Intelligence (SI) is the property of a system whereby the collective behaviors of (unsophisticated) entities interacting locally with their environment cause coherent functional global patterns to emerge. SI provides a basis with which it is possible to explore collective (or distributed) problem solving without centralized control or the provision of a global model (Stan Franklin, *Coordination without Communication*, talk at Memphis Univ., USA, 1996). The well-know bio-inspired computational paradigms know as ACO (*Ant Colony Optimization* algorithm [4]) based on trail formation via pheromone deposition / evaporation, and PSO (*Particle Swarm Optimization* [9]) are just two among many successful examples. To tackle the formation of a coherent social collective intelligence from individual behaviors, in the present work, we will address the collective adaptation of a social community to a cultural (environmental, contextual) or informational dynamical landscape, represented here – for the purpose of different experiments – by several three-dimensional mathematical functions that change over time. Also, unlike past works [13], the size of our swarm population varies over time: agents reproduce and die, according to its success on reaching the peaks (e.g., if the swarm is requested to find the higher values of one complex function) or deep valleys (if the goal is minimization) over the three-dimensional landscapes. We believe that *Swarms with Varying Population Size* (SVPS) provide a better model to mimic some natural features, improving not only the population ability to evolve self-organized foraging behavior as obtained in the past [13], while maintaining a self-regulated population adapted in real-time to different constraints in different search landscapes. The complexity of the system, not only provides more efficiency on the task of finding peaks/valleys, as well as supplies a faster response to the changing environment (changing the search landscape during the run). The progress of the population size over time also suggests that our system may be evolving in a self-organized critical state [3], surpassing several phase transitions. The present work is organized as follows; Section 2 gives an overview of related work in the area of artificial life models with varying population size. Section 3 describes the original swarm model, before the dynamic of population variation was introduced. The proposed swarm model and its properties are described in section 4. In Section 5 the results are shown and its implications are discussed. Finally, Section 6 concludes the paper and suggests future research, namely in what relates to Co-Evolution.

## 2 Related work

In [2,3], Bak modeled an ecological system, consisting of different species represented by random fitness values, each interacting with two neighbors. After mutating, at each time step, the least fit species, they measured the size of mutation avalanches on the system. They concluded that the phenomenon follows a power law with slope approximately equal to 1, suggesting that the system self-organizes to a critical state.

Genetic Algorithms (GAs) are usually implemented using populations with fixed size. However seeking for extra performing systems, in [1], the GAVaPS - Genetic Algorithm with Varying Population Size - was presented, and many works in this area have been made since then. In this GA, the concept of "age" of an individual is introduced. A chromosome remains in the population (i.e. stays "alive") for a number of generations proportional to its fitness (*lifetime*). When the age of a chromosome, which is incremented each generation, reaches its lifetime, the individual is removed from the population. Consequently, there is no direct selection pressure in this GA. The individuals are randomly selected for crossover and mutation. The pressure is assured by the fact that fitter individuals remain in the population during a larger number of generations, thus producing more offspring than those with lower fitness. The authors tested the algorithm with four different functions and concluded that it outperforms the Simple GA in some cases. They also showed that the population size, after a large initial growth, decreases, and remains stable around the initial value. Although the results seem promising, the test functions were not very demanding, and the GAs, simple and with varying population size, attained near-optimal solutions in few function evaluations (between 1000 and 2000). The GAVaPS was tested again in [7] by Fernandes *et al.* with the Royal Road function R4. The R4 problem has different characteristics than those previously tested and demands a higher computation.

The population behavior described above was not observed in the tests made with the Royal Road function: the population either decreased towards mass extinction, or dramatically increased to enormous dimensions. These results exposed some weak points of GAVaPS, an algorithm that appears to evolve away from the self-organized criticality state. In order to achieve a self-organized behavior it is essential to review the reproduction strategy of GAVaPS. The work presented in this paper may also lead to an insight into the proper directions to follow when creating Evolutionary Algorithms with varying population size.

## 3 The Swarm Landscape Foraging Model

The swarm intelligence algorithm fully uses agents that stochastically move around the “habitat” following pheromone concentrations. That is, instead of trying to solve minimization or maximization problems by adding different ant casts, short-term memories and behavioral switches, which are computationally intensive, representing simultaneously a potential and difficult complex parameter tuning, it was our intention to follow real ant-like foraging behaviors.

$$W(\sigma) = \left(1 + \frac{\sigma}{1 + \delta\sigma}\right)^{\beta} \quad (1) \qquad P_{ik} = \frac{W(\sigma_i)w(\Delta_i)}{\sum_{j/k} W(\sigma_j)w(\Delta_j)} \quad (2) \qquad T = \eta + p\frac{\Delta[i]}{\Delta_{\max}} \quad (3)$$

In that sense, bio-inspired spatial transition probabilities are incorporated into the system, avoiding randomly moving agents, which tend the distributed algorithm to explore regions manifestly without interest (e.g., valleys, when the goal is to search peaks on fitness landscapes), being generally, this type of exploration, counterproductive and time consuming. Since this type of transition probabilities depend on the spatial distribution of pheromone across the environment, the behavior reproduced is also a stigmergic one [6,12]. Moreover, the strategy not only allows guiding ants to find peaks/valleys in an adaptive way, as the use of embodied short-term memories is avoided. As we shall see, the distribution of the pheromone represents the memory of the recent history of the swarm, and in a sense it contains information which the individual ants are unable to hold or transmit. There is no direct communication between the organisms but a type of indirect communication through the pheromonal field. In fact, ants are not allowed to have any memory and the individual’s spatial knowledge is restricted to local information about the whole colony pheromone density. In order to design this behavior, one simple model was adopted [5], and extended (as in [12]) due to specific constraints of the present proposal. As described in [4], the state of an individual ant can be expressed by its position $r$, and orientation $\theta$. It is then sufficient to specify a transition probability from one place and orientation $(r,\theta)$ to the next $(r^*,\theta^*)$ an instant later. The response function can effectively be translated into a two-parameter transition rule between the cells by use of a pheromone weighting function (Eq. 1). This equation measures the relative probabilities of moving to a cite $r$ (in our context, to a grid location) with pheromone density $\sigma(r)$. The parameter $\beta$ is associated with the osmotropotaxic sensitivity (a kind of instantaneous pheromonal gradient following), and on the other hand, $1/\delta$ is the sensory capacity, which describes the fact that each ant's ability to sense pheromone decreases somewhat at high concentrations. In addition to the former equation, there is a weighting factor $w(\Delta\theta)$, where $\Delta\theta$ is the change in direction at each time step, i.e. measures the magnitude of the difference in orientation. As an additional condition, each individual leaves a constant amount $\eta$ of pheromone at the cell in which it is located at every time step $t$. This pheromone decays at each time step at a rate $k$. Then, the normalized transition probabilities on the lattice to go from cell $k$ to cell $i$ are given by $P_{ik}$ [5] (Eq. 2), where the notation $j/k$ indicates the sum over all the pixels $j$ which are in the local neighborhood of $k$. Finally, $\Delta_i$ measures the magnitude of the difference in orientation for the previous direction at time $t$-1.

In order to achieve emergent and *autocatalytic* mass behaviours around specific locations (e.g., peaks or valleys) on the *habitat*, instead of a constant pheromone deposition rate $\eta$ used in [5], a term not constant is included. This upgrade can significantly change the expected ant colony cognitive map (pheromonal field). The strategy follows an idea implemented earlier by Ramos in [12], while extending the Chialvo model into clustering purposes, aiming to achieve a collective perception of those images by the end product of swarm interactions. The main differences to the Chialvo work, is that ants now move on a 3D discrete grid, representing the functions which we aim to study instead of a 2D *habitat*, and the pheromone update takes in account not only the local pheromone distribution as well as some characteristics of the cells around one ant. In here, this additional term should naturally be related with specific characteristics of cells around one ant, like their altitude ($z$ value or function value at coordinates $x,y$), due to our present aim. So, our pheromone deposition rate $T$, for a specific ant, at one specific cell $i$ (at time $t$), should change to a dynamic value ($p$ is a constant = 1.93) expressed by equation 3. In this equation, $\Delta_{max} = | z_{max} - z_{min} |$, being $z_{max}$ the maximum altitude found by the colony so far on the function *habitat*, and $z_{min}$ the lowest altitude. The other term $\Delta[i]$ is equivalent to (if our aim is to minimize any given landscape): $\Delta[i] = | z_i - z_{max} |$, being $z_i$ the current altitude of one ant at cell $i$. If on the contrary, our aim is to maximize any given landscape, then we should instead use $\Delta[i] = | z_i - z_{min} |$. Finally, notice that if our landscape is completely flat, results expected by this extended model will be equal to those found by Chialvo and Millonas in [5], since $\Delta[i]/\Delta_{max}$ equals to zero. In this case, this is equivalent to say that only the swarm pheromonal field is affecting each ant choices, and not the *environment* - i.e. the expected network of trails depends largely on the initial random position of the colony, and in trail clusters formed in the initial configurations of pheromone. On the other hand, if this environmental term is added a stable and emergent configuration will appear which is largely independent on the initial conditions of the colony and becomes more and more dependent on the nature of the current studied *landscape* itself. The environment plays an active role, in conjunction with continuous positive and negative feedbacks provided by the colony and their pheromone, in order to achieve a stable emergent pattern, memory and distributed learning by the community [13].

# 4 The Swarm Model with Varying Population Size

The model described in section 3 is a *Swarm with Fixed Population Size* (SFPS). In this paper we propose and aim to analyze the behavior of a *Swarm with Varying Population Size* (SVPS). This characteristic is achieved by allowing ants to reproduce and die through their evolution in the fitness landscapes. To be effective, the process of variation must incorporate some kind of pressure towards successful behavior, that is, ants that reach peaks/valleys must have some kind of reproductive reward, by staying alive for more generations – generating more offspring - or by simply having a higher probability of generating offspring in each time step. In addition, the population density in the area surrounding the parents must be taken into account when a reproduction event is about to occur. By having these issues in mind, we developed a simple model of population variation with two rules that control the aging (which inevitably leads to death) and reproduction of the agents.

## 4.1 Aging Process / Reproduction Process

When one ant is created (during initialization or by the reproduction process) a fixed energy value is assigned to it. Every time step, the ant energy is decreased by a constant amount. The ant's probability of survival after a time step is proportional to its energy during that same iteration. In the model used in the tests shown below we normalized the energy, by setting its initial value to 1 - meaning that the probability of surviving the first iteration equals 1. Now, the energy of each ant is decreased by 0.1 on every iteration, which means that after ten generations (occurring on the whole population), this and other ants will inevitably die. Within these settings one ant that is for instance, 7 iterations old, has a probability of 0.3 to survive at the current time step. Meanwhile, for the reproduction process, we assume the following heuristic: an ant (main parent) triggers a reproduction procedure if it finds at least another ant occupying one of its 8 surrounding cells (*Moore* neighborhood is adopted). The probability $P(r)$ of generating offspring – one child for each reproduction event – is computed in two steps. First, the surrounding area is inspected in order to see if it is too crowded. Being $n$ the number of occupied cells around this ant, the probability to reproduce is set to the values shown in table 1 ($P^*$). Notice that: 1) an ant completely surrounded by other ants, or isolated ($n$=8, $n$=0) do not reproduce; 2) the maximum probability is achieved when the area around that ant is half occupied ($n$=4).

$$P(r) = P^* \frac{\Delta[i]}{\Delta_{\max}} \qquad (4)$$

**Table 1.** Reproduction probability $P^*$ values according to the number of *Moore* neighbors

| ***n*=0** or ***n*=8** | ***n*=4** | ***n*=5** or ***n*=3** | ***n*=6** or ***n*=2** | ***n*=7** or ***n*=1** |
|---|---|---|---|---|
| 0 | 1 | 0.75 | 0.5 | 0.25 |

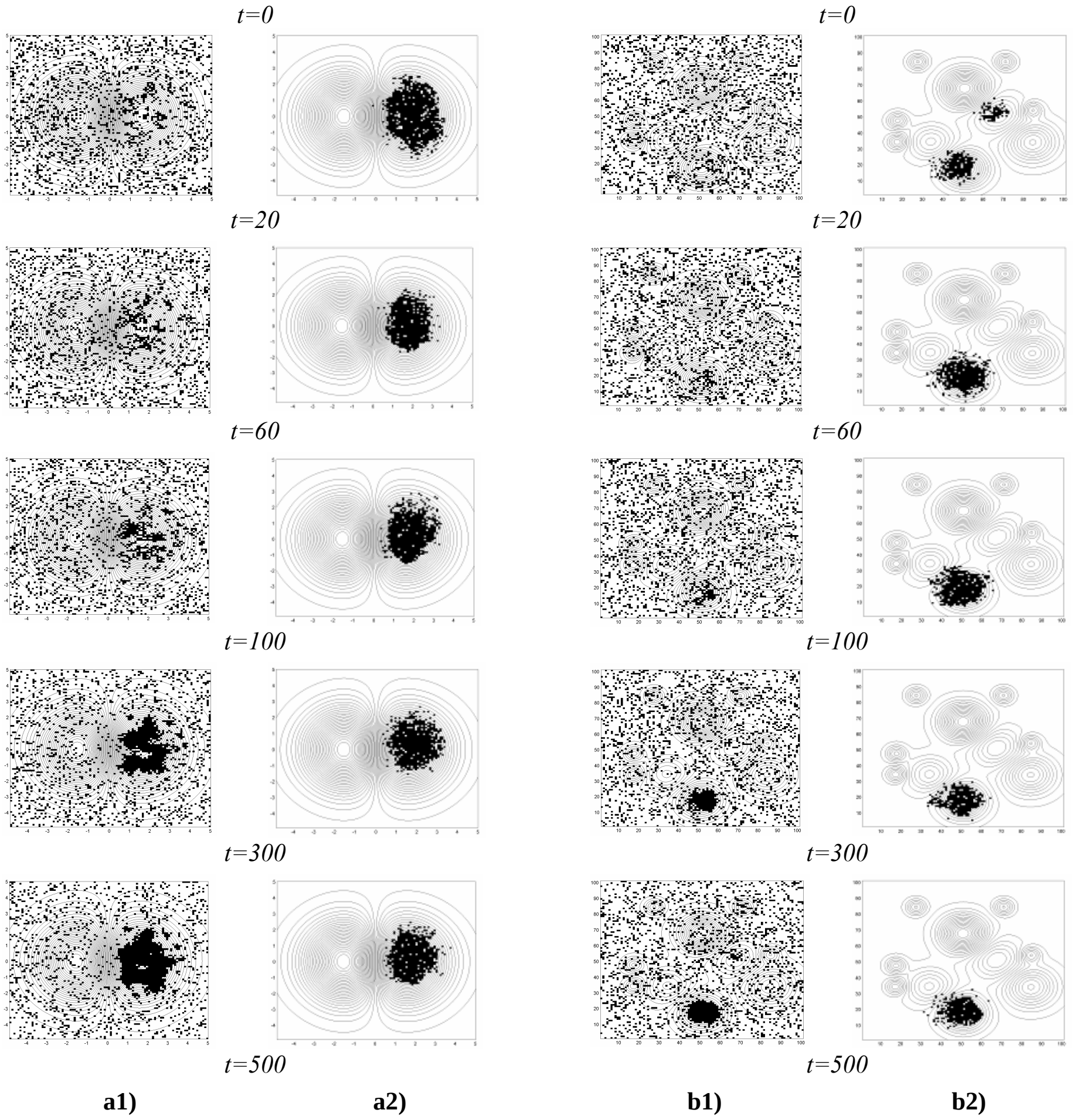

**Figure 1.** Evolution of the swarm during 500 iterations for a) maximization of *F0a*, and b) minimization of *Passino F1* (check fig. 2). The figures in columns a1) and b1) correspond to the ants distribution in the habitat in a SFPS while a2) and b2) are the result of the SVPS. (Parameters: a1): $\beta = 7$, *size = 20%*; a2): $\beta = 7$, *initial size = 10%*; b1): $\beta = 3.5$, *size = 20%*; b2): $\beta = 3.5$, *initial size = 10%*).

After the probability $P^*$ is set to one of the previous values, the final probability is computed according to Eq. 4, with $\Delta[i]$ and $\Delta_{max}$ as in Eq. 3. This operation guarantees that any ant reaching the higher(lower) peaks(valleys) have more chance to produce offspring (notice that one ant in the higher/lower cell has $P(r)$=1 if $n$=4 and will repro-

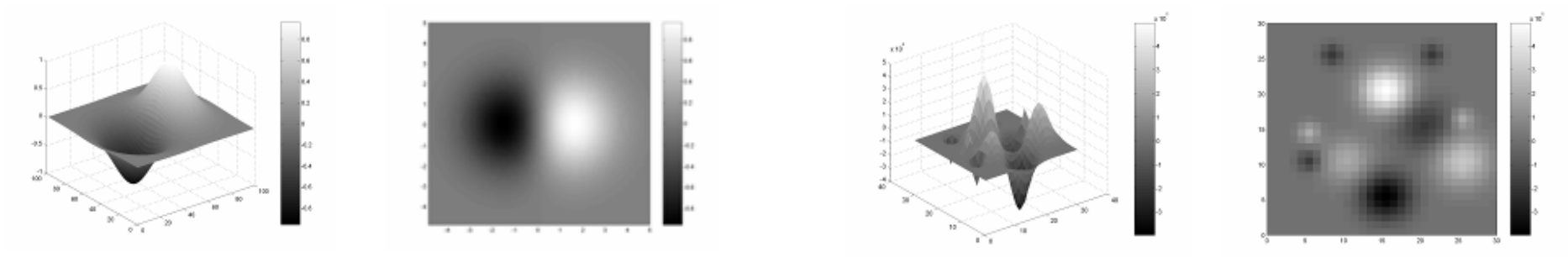

*F0a - 3D view* *F0a - 2D view* *Passino F1 – 3D* *Passino F1 – 2D*

**Figure 2.** The 2D and 3D representation of *F0a* and *Passino F1* functions (in the 2D graphics the darker spots represent the deeper valleys).

duce for certain). If this ant passes the reproduction test a new agent is created, occupying one of his vacant cells around the main parent. Any infant ants are allowed to be allocated in places where other ants are.

## 5 Results

From the set of functions used in [13] to test the fixed population size swarm, we chose *F0a*, *F6* and *Passino F1* (fig. 2) to compare the performance of the two models. The functions were adapted to a 100 x 100 toroidal grid (see figure 1 and 2). Parameters $\sigma, \eta, k$ were set to 0.2, 0.07 and 1.0, respectively, following previous tests conducted in [5] and [13]. We kept these values constant during the tests in order to keep the analysis simple and tested several configurations over SFPS and SVPS with $\beta$ between 1 and 15 and swarm size (initial size, in the SVPS case) between 5 and 50 percent of the habitat size (10.000 cells resulting from the 100 x 100 grid). By observing the results we concluded that SVPS clearly outperforms the previous fixed sized swarm algorithm when maximizing and minimizing the test functions, as well as the *Bacterial Foraging Optimization Algorithm*, BFOA, presented earlier by Passino *et al.* [11,10], also compared in [13]. In figure 1 we can see that SVPS converges much faster to the desired regions of the habitat. We also notice that the variation of population mechanism eliminates the wandering ants and drives the entire swarm towards peaks or valleys, simultaneously self-regulating the population when need it (e.g., peaks with a small area). This behavior, however, does not affect the capability of the swarm to adapt to a changing environment, as we shall see below. Fig. 3 describes the evolution of the median height of the swarm for several configurations of the algorithm applied to the maximization of *F0* and minimization of *Passino F1*. Comparing the curves originated by SFPS and SVPS (as well as BFOA, [13,11,10]) it is clear that the model with varying population size adapts better to the landscapes and conducts the swarm to the objective in a faster and more accurate way.

### 5.1 The Importance of the Self-Organizing process via Pheromonal cues

Such a clear difference between the performances of the two models may lead to the wrong idea that the population dynamics (*Reproduction*) is the sole responsible for the good performance of the SVPS and that pheromonal fields (*Self-Organization*, [13])

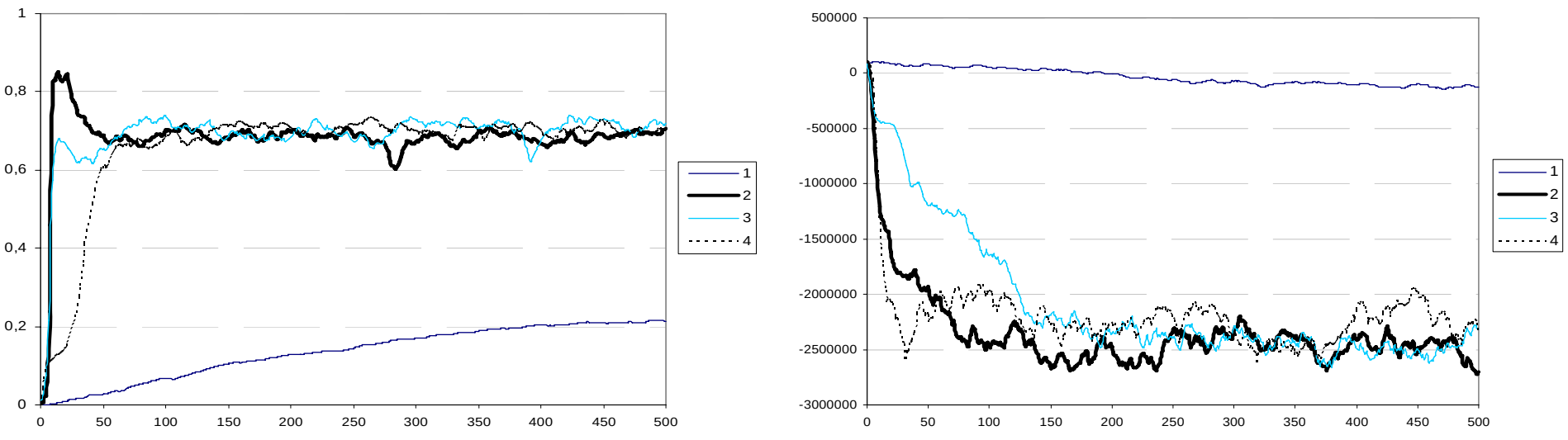

*a)* *b)*

**Figure 3.** Median height of the ants during swarm evolution when maximizing F0 (a) and minimizing Passino F1 (b). Parameters: **a)** 1-SFPS: β=7, size=20%; 2-SVPS: β=3.5; initial size=10%; 3- SVPS: β=7; initial size=5%; 4- SVPS: β=3.5; initial size=30%. **b)** 1-SFPS: β=3.5, size=20%; 2-SVPS: β=15; initial size=10%; 3- SVPS: β=7; initial size=20%; 4- SVPS: β=3.5; initial size=10%.

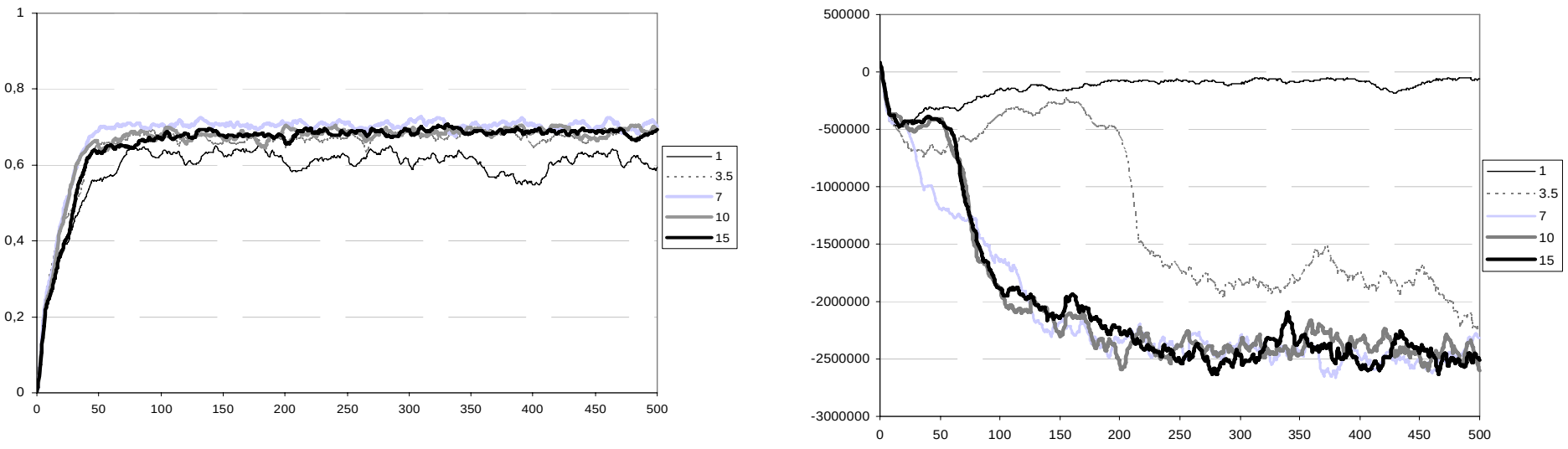

**Figure 4.** Median height of the ants during swarm evolution when maximizing *F0* (a) and minimizing *Passino F1* (b). The curves refer to SVPS with *β* equal to 1, 3.5, 7 and 15. Both **a)** and **b)** populations have initial size equal to 20 percent of the search space in all configurations.

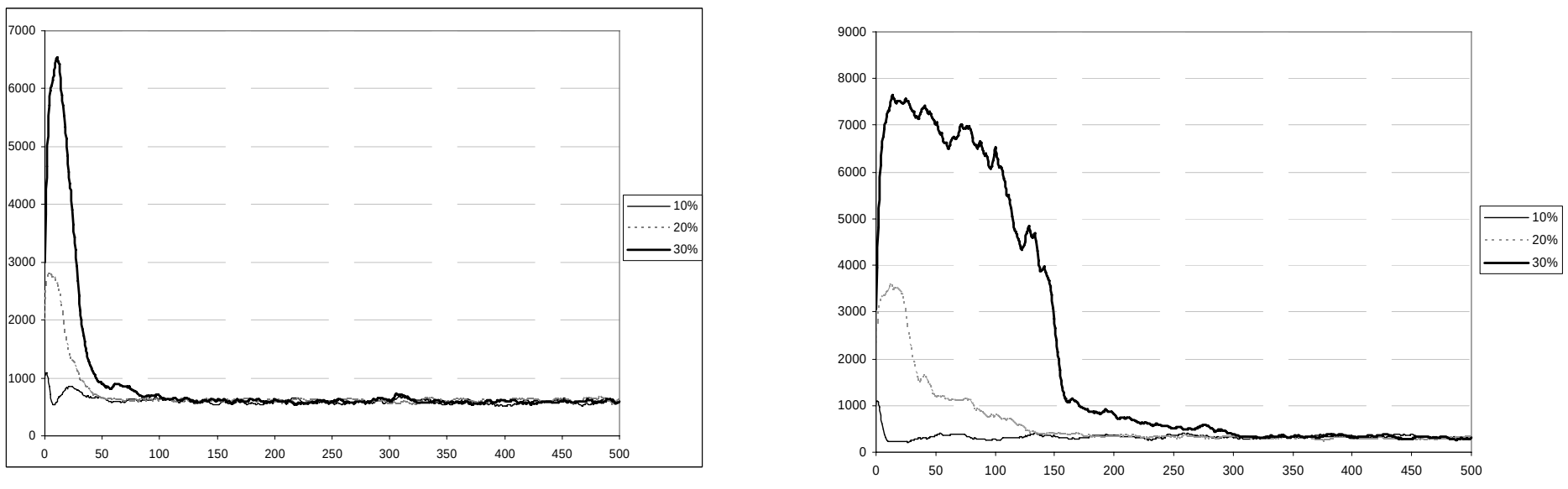

*a)* *b)*

**Figure 5.** SVPS maximizing *F0* and minimizing *Passino F1* with initial population size equal to 10, 20 and 30 percent of the search space. All runs were made using *β = 7*.

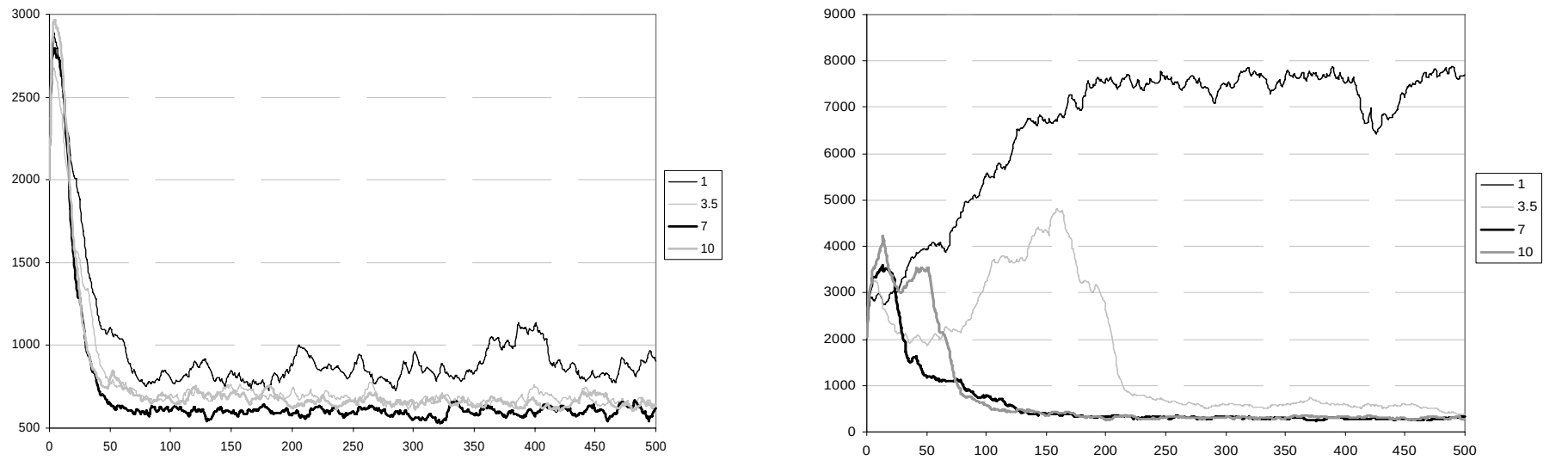

***a)*** ***b)***

**Figure 6.** Population growth of the SVPS when a) maximizing *F0* and b) minimizing *Passino F1*. All runs were made with an initial population size equal to 20 percent of the search space.

are playing a back role. To investigate this possibility we tested SVPS with different values of $\beta$, from 1 (corresponding to a very low or absent tendency to follow pheromone) to 15. By analyzing the evolution of the median height of the swarm on the landscape we concluded that pheromone following is essential to a fast and self-organized convergence to peaks/valleys. Figure 4 shows the median height of our swarm over several time steps, for the maximization of *F0* and the minimization of *Passino F1* [13,11,10]. In the first case the difference is not so obvious although a closer inspection reveals the poorer performance of $\beta=1$ configuration compared to swarms with a stronger impulse to follow pheromone. Meanwhile the *Passino F1*, with its multiple peaks and valleys, creates more problems to the colony and truly reveals the utility of the positive (*pheromone reinforcement* = *exploitation behavior*) and negative feedbacks (*evaporation* = *explorative behavior*) introduced by the pheromonal fields [13]. It is clear, in this last case, that $\beta=1$ doesn't lead to a converging algorithm and even a swarm with $\beta=3.5$ experiences some problems on climbing the higher peak of the function. These results show that although pheromone following and varying population mechanisms can lead the swarms to the desired regions, the cooperation between the two processes result in a much powerful system.

### 5.2 Population Size over Time

By plotting the evolution of the population size of SVPS, we concluded that for each function and task (minimization/maximization) the population tends to evolve until it stabilizes around a specific value. In figure 5 we can see this behavior when maximizing *F0* and minimizing *Passino F1*, as $\beta$ remains fixed and the initial population size varies from 10% to 30% of the habitat size. Notice that the final values of the population size differ for the two functions: within *F0* the population becomes stable around 600 agents while the swarms minimizing *Passino F1* evolved populations with near 300 individuals. If we try to maximize *Passino F1* or evolve the swarm into other functions we see that the populations become stable around different values. That is, the mechanism here introduced is able to self-regulate the population according to the actual foraging environment (swarms adapt to it). So much so, that a similar result could be obtained over dynamic landscapes (fig. 7) as we shall see later (section 5.3). This pattern however, is broken when the value of $\beta$ decreases (representing a less self-organized behavior). As we can see in figure 6, $\beta$ equal to 1 leads the swarm to an unpredictable and even rather chaotic behavior (figure 6b). The limits to the population growth are imposed by the search space itself because only one ant may occupy each cell. Under these conditions, it is impossible to observe exponential growing of population size. On the other hand, mass extinction is possible. Almost all the extinctions observed in the exhaustive SVPS testing were related to models with $\beta$=1 (although in a few configurations the extinction phenomenon was observed at $\beta$=3.5). These results emphasize the role of Self-Organization while evolution occur trough any requested aim (our task), enhancing the increasing recognition that *Natural Selection* (here coded via the reproduction process) and *Self-Organization* (here coded via the pheromone laying process) work hand in hand to form *Evolution*, as defended by Kauffmann [8].

### 5.3 Time Varying Landscape Distributed Search

One of the features of SFPS discussed in [13] was the ability to adapt to sudden changes in the roughness of the landscape. These changes were simulated by abruptly replacing one test function by another after the swarm reaches the desired regions of the landscape. Another way of simulating changes in the environment consists on changing the task from minimization to maximization (or vice-versa). The swarm performance was convincing and reinforced the idea that the system is highly adaptable and flexible. In here, we followed and conducted similar tests using SVPS and concluded that varying population size increases the capability of the swarm to react to changing landscapes. Figure 7 shows SFPS (6a) and SVPS (6b) trying to find the lower values of *Passino F1* until $t$=250, and then searching for the higher values (minimization-maximization of the function). We can see that SVPS not only evolves faster to valleys as it readapts better to the changing goal. In fact, at $t$=500, SFPS is still migrating to the higher peak in the landscape while SVPS has already performed the task at $t$=340. Notice, by observing the 2D and 3D representation of the function (figure 2) and the maps in figure 7, how SVPS uses the landscape specific characteristics to emerge a wonderful and distinctive migration process (as a *flocking behavior*) after $t$=250: the swarm climbs the neighboring peak (local optima) on the left and from that point (forming a explorative *Mickey Mouse* shape at $t$=300) finally reaches the higher peak of the landscape, where it remains while self-regulates the population; by some complex mechanism the swarm does not climb the peak on the right side of the valley (which is higher) avoiding the valley that stands between that local optima and the desired region (see the 2D representation of *Passino F1*).

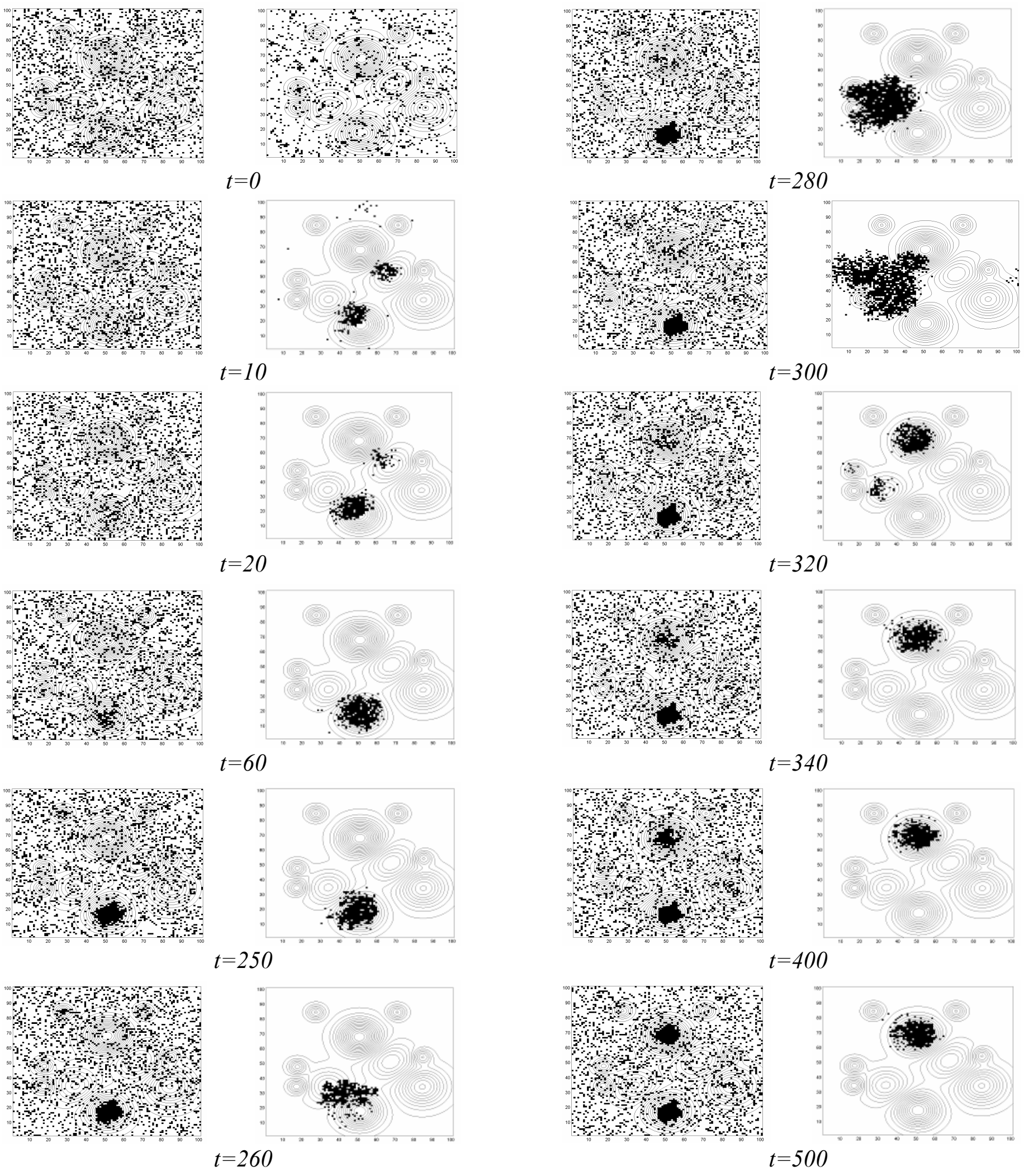

**Figure 7.** SFPS (left) and SVPS(right) evolving in Passino F1 function. From *t=0* to *t=250* the swarm is induced to search the valleys of the landscape. After *t=250* the task changes and the swarm must find the higher values of the function.

## 6 Conclusions and Future Work

We showed that adding a mechanism of varying population size to a self-organized swarm search algorithm increases not only its capability to search/find peaks and valleys of fitness landscapes, but it also provides the system with some unexpected self-regulated behavior – for instance, population growth to predictable values even with rather different $\beta$ and initial population size values. SVPS also proved to be more effective when evolving over changing dynamic landscapes. The way the popu-lation grows during the process and how it stabilizes around certain values requires insight into the self-organization of the system. It is crucial to check whether the swarm evolves to a self-organized critical state or if it is not robust enough to be classified in that way. It is also important to understand the reason why the population size evolves to different values over different functions and if there is a way to predict that value by observing the roughness of the landscape.

The connection between the SVPS and co-evolutionary artificial life models is also worth inspecting since species have to adapt to constant changes in the fitness landscape caused by mutations in other species. The ability revealed by the SVPS to quickly readapt to changing environments suggests its utility in the study of co-evolutionary systems dynamics.